\documentstyle[epsf,floats,prb,aps]{revtex}

\begin{document}
\draft

\twocolumn[\hsize\textwidth\columnwidth\hsize\csname
@twocolumnfalse\endcsname

\title{Heterovalent and A-atom effects in A(B$'$B$''$)O$_{3}$ perovskite alloys}
\author{L. Bellaiche, J. Padilla and David Vanderbilt}
\address{Department of Physics and Astronomy,\\
         Rutgers University, Piscataway, New Jersey 08855-0849, USA}
\date{August 24, 1998}
\maketitle

\begin{abstract}
Using first-principles supercell calculations, we have investigated
energetic, structural and dielectric properties of three
different A(B$'$B$''$)O$_{3}$ perovskite alloys:
Ba(Zn$_{1/3}$Nb$_{2/3}$)O$_{3}$ (BZN),
Pb(Zn$_{1/3}$Nb$_{2/3}$)O$_{3}$ (PZN), and
Pb(Zr$_{1/3}$Ti$_{2/3}$)O$_{3}$ (PZT).
In the homovalent alloy PZT, the energetics are found to be mainly
driven by atomic relaxations.  In the heterovalent alloys BZN and PZN,
however, electrostatic interactions among B$'$ and B$''$ atoms
are found to be very important.
These electrostatic interactions are responsible for the stabilization
of the observed compositional long-range order in BZN.  On the
other hand, cell relaxations and the formation of short Pb--O bonds
could lead to a destabilization of the same ordered structure in
PZN.
Finally, comparing the dielectric properties of homovalent and
heterovalent alloys, the most dramatic difference arises in
connection with the effective charges of the B$'$ atom.  We
find that the effective charge of Zr in PZT is anomalous, while
in BZN and PZN the effective charge of Zn is close to its nominal
ionic value.
\end{abstract}


\vskip2pc]

\narrowtext

\section{Introduction}

Simple perovskite compounds have the chemical formula ABO$_{3}$.
In the cubic phase, the oxygen atoms form a cubic lattice of
corner-sharing octahedra with the B cations at their centers,
while the A cations form a second interpenetrating cubic sublattice
located at the 12-fold coordinated sites between octahedra.
Interestingly, most of the perovskite compounds that are of greatest
technological interests are not simple systems,
but rather A(B$'$B$''$)O$_{3}$ alloys with two different kinds
of B atoms.
Examples include the Pb(ZrTi)O$_{3}$ alloys that are
currently used in
piezoelectric transducers and actuators
\cite{Uchino,Tanaka,Sheppard}.
Other examples include
Ba(MgNb)O$_{3}$ for high-frequency applications \cite{Akbas2}, and
Pb(MgNb)O$_{3}$ and Pb(ZnNb)O$_{3}$ which exhibit extraordinarily
high values of the piezoelectric constants when alloyed with
PbTiO$_{3}$ \cite{Park}.

From a chemical point of view, two classes of A(B$'$B$''$)O$_{3}$
compounds can be distinguished: homovalent {\it vs.}\ heterovalent
alloys.  In homovalent alloys, the two B atoms belong to the same
column of the periodic table. The A(B$'_{x}$B$''_{1-x}$)O$_{3}$
alloys can thus have a composition $x$ continuously ranging from 0 to
1.  A typical example of such a homovalent alloy is
Pb(Zr$_{x}$Ti$_{1-x}$)O$_{3}$.  On the other hand, there is a unique
stoichiometry in heterovalent alloys, since the two B atoms do not
belong to the same column of the periodic table.  Typical examples
of heterovalent alloys are
Pb(Zn$_{1/3}$Nb$_{2/3}$)O$_{3}$,
Ba(Zn$_{1/3}$Nb$_{2/3}$)O$_{3}$,
Pb(Mg$_{1/3}$Nb$_{2/3}$)O$_{3}$,
Ba(Mg$_{1/3}$Nb$_{2/3}$)O$_{3}$,  and
Pb(Sc$_{1/2}$Ta$_{1/2}$)O$_{3}$.

While many first-principles calculations have been performed to
understand and predict properties of simple ABO$_3$ perovskite
systems (see Ref.~\cite{Davidreview,SzaboPRL,Nicola} and references
therein), only a few studies
have investigated alloy properties via accurate {\it ab-initio}
techniques.  For example, ferroelectric effects in PZT have been
studied \cite{LaurentJorgeDavid}, and energetics of long-range
order structures of homovalent or heterovalent alloys have been
investigated \cite{Szabo,Wensell,Burton1,Burton2}.  These
recent studies provide a useful understanding of the microscopic
behavior of perovskite alloys, but they do not give a simple picture
of the consequences of heterovalency on the energetics and properties
of the alloys.  For example, one may wish to compare the driving
mechanisms responsible for the energetics of homovalent and
heterovalent alloys. It is also unclear how the chemical difference
between homovalent and heterovalent alloys affects interatomic
distances and dielectric properties.

Another aspect that is poorly understood is the effect of the A-atom
identity on properties of heterovalent alloys. For example, many
Ba-compounds including
Ba(Mg$_{1/3}$Nb$_{2/3}$)O$_{3}$ (BMN) \cite{Akbas2},
Ba(Zn$_{1/3}$Nb$_{2/3}$)O$_{3}$ (BZN) \cite{Treiber},
Ba(Zn$_{1/3}$Ta$_{2/3}$)O$_{3}$ (BZT) \cite{Jacobson,Allen},
Ba(Mg$_{1/3}$Ta$_{2/3}$)O$_{3}$ (BMT) \cite{Guo},
Ba(Sr$_{1/3}$Ta$_{2/3}$)O$_{3}$ (BST) \cite{Guo} and
Ba(Ni$_{1/3}$Nb$_{2/3}$)O$_{3}$ (BNN) \cite{Kim},
adopt
a structure with long-range compositional order in which one plane
of B$'$ atoms alternates with two planes of B$''$ atoms along
the [111] direction.  We shall refer to this as the 1:2$_{[111]}$
structure.  On the other hand, weak X-ray reflections have been
detected in various Pb-compounds, such as Pb(Mg$_{1/3}$Nb$_{2/3}$)O$_{3}$
(PMN) \cite{Husson1,Husson2,Chen1} and Pb(Mg$_{1/3}$Ta$_{2/3}$)O$_{3}$
(PMT) \cite{Akbas1}, that are indicative of a rocksalt-like 1:1
ordering of B$'$-rich and B$''$-rich sites on the B sublattice.  We
will refer to this as the 1:1$_{[111]}$ structure, since it can be
regarded as consisting of a simple alternation of planes of B$'$-rich
and B$''$-rich sites along [111].  The nature and strength of the
atomic ordering have been shown to be crucial for the properties
of this class of alloys.

Understanding heterovalent and A-atom effects can also be very
useful for building an accurate effective Hamiltonian that can
extend the reach of first-principles calculations by retaining only
the physically relevant degrees of freedom \cite{Wensell}.
Effective Hamiltonian approaches have been developed in the past
for simple perovskite systems. When constructed from first-principles
calculations, they have proven to be very successful both for reproducing
phases transition sequences \cite{ZhongDavid,Rabe,Krakaeur}, and
for studying ferroelectric
domain walls \cite{Jorge} as well as finite-temperature dielectric and
electromechanical properties \cite{Alberto1,Alberto2,Cockayne}.
An effective Hamiltonian for a perovskite alloy should include both
compositional and structural degrees of freedom.  A good starting
point for treating the former may be the simple model of
Ref.~\cite{LaurentDavid}.  By focusing on electrostatic effects,
this model successfully reproduces the compositional long-range
order observed in a large class of complex heterovalent perovskite
alloys.  However, in order to augment this model with a description
of structural relaxations, we are motivated to understand the role
of such relaxations in the energetics and the dielectric properties
of these materials.

In the present study, we investigate both heterovalent and A-atom
effects in an important class of perovskite alloys using
first-principles techniques applied to ordered supercells.
To study heterovalent effects, we compare the computed properties
of Pb(B$'$B$''$)O$_{3}$  alloys for the cases of homovalent
vs.\ heterovalent B-site substitution.
With regard to A-atom effects, we are mainly interested in the
differences between Pb compounds and other divalent-A compounds.
While we find that the
energetics of homovalent alloys are mainly driven by atomic relaxations,
we confirm that the electrostatic interactions are extremely
important to understand the energetic and other properties of heterovalent
alloys.
For example, electrostatic interactions are mainly responsible for the
long-range order seen in heterovalent Ba-compounds.
In addition, we find that the
formation of short Pb--O bonds in heterovalent
Pb(B$'$B$''$)O$_{3}$  alloys could help in destabilizing the
structure suggested by electrostatic considerations alone.
Finally, the present study also emphasizes that the dielectric properties
of homovalent and heterovalent alloys can differ considerably. For
example, the effective charge of the B$'$ atom is found to
be anomalous in homovalent alloys, while it is close to its nominal
value in heterovalent alloys.

Specifically, we have chosen three compounds for study.  These are
the homovalent Pb(Zr$_{x}$Ti$_{1-x}$)O$_{3}$ (PZT) system
and the heterovalent Pb(Zn$_{1/3}$Nb$_{2/3}$)O$_{3}$ (PZN) and
Ba(Zn$_{1/3}$Nb$_{2/3}$)O$_{3}$ (BZN) systems.
To facilitate direct comparison between the homovalent and heterovalent
alloys, we choose the same composition $x=1/3$ for PZT as for PZN and BZN.
Note that in all three alloys the larger B atom (Zn or Zr) is the minority
species, while the smaller one (Nb or Ti) is the majority.

All the supercells studied in the present article retain
inversion symmetry, thus preventing the occurrence of
ferroelectricity and piezoelectricity.  While the Pb compounds
PZN and PZT do exhibit ferroelectricity at low temperature,
many interesting features (such as short Pb--O bonds \cite{Egami})
appear in the paraelectric phase of these compounds.  Moreover,
the atomic ordering on the B sublattice, which is a focus of
the present work, is determined at growth temperatures well
above the ferroelectric-to-paraelectric transition temperature.
For the purpose of understanding the
intrinsic differences between homovalent and
heterovalent alloys, and between
heterovalent alloys with two different A atoms, we thus decided not to relax
the inversion-symmetry constraint.

\section{Methods}

We limit ourselves to the two ordered superlattice structures
shown in Fig.~1.  These 15-atoms supercells consist of one B$'$
\begin{figure}
\epsfxsize=3.0in
\centerline{\epsfbox{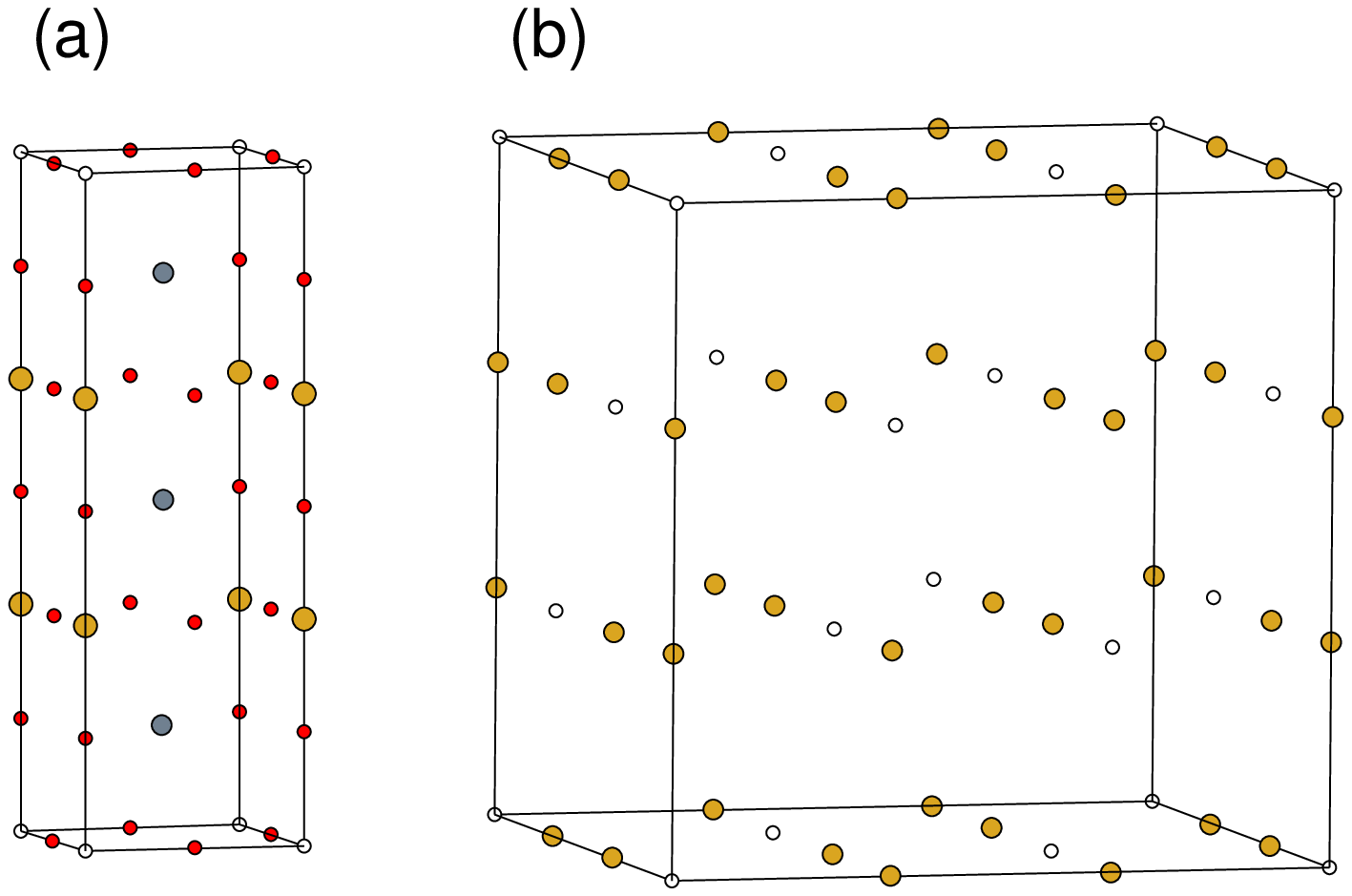}}
\vskip 0.2truein
\caption{
(a) The 1:2$_{[001]}$ A$_{3}$(B$'$B$''$$_{2}$)O$_{9}$ supercell.
The succession of planes from bottom to top (i.e., progressing
along [001]) is
(B$'$,O), (Pb,O), (B$''$,O), (Pb,O), (B$''$,O), (Pb,O)...
(b) B-site atomic ordering in the 1:2$_{[111]}$ structure,
illustrated within a large box aligned with the cartesian axes.
The succession of planes progressing along 
the [111] direction is
(B$'$), (Pb,O), (B$''$), (Pb,O), (B$''$), (Pb,O)...;
only B$'$ and B$''$ atoms (small open and large
filled circles respectively) are shown here.
}
\end{figure}
plane alternating with two B$''$ planes along either the
[001] direction (Fig.~1a) or the [111] direction (Fig.~1b).
The structure shown in Fig.~1a will be denoted as 1:2$_{[001]}$.
As previously indicated, 1:2$_{[111]}$ denotes the structure shown in Fig.~1b.
The lattice vectors of the
1:2$_{[001]}$ structure are
${\bf a}_{[001]}=a_0[1,0,0]$,
${\bf b}_{[001]}=a_0[0,1,0]$,
and ${\bf c}_{[001]}=\xi a_0[0,0,3]$,
where $a_0$ is the lattice parameter and $\xi$ is the ordered-related
axial ratio. The ideal
1:2$_{[001]}$ structure exhibits the tetragonal
4/mmm point group.
On the other hand, the point group of the ideal
1:2$_{[111]}$ structure is
$\bar{3}$m, and its lattice vectors are
${\bf a}_{[111]}=a_0[1,0,-1]$,
${\bf b}_{[111]}=a_0[-1,1,0]$,
and ${\bf c}_{[111]}=\xi a_0[1,1,1]$, where $a_0$ is
the lattice parameter and $\xi$ is the ordered-related
axial ratio.
We will retain the ideal group symmetry in both structures.
The two B$''$ atoms are then equivalent by symmetry in
both structures.  Of the three A atoms in the cell,
the two A atoms located between a B$'$ and a B$''$ plane
are equivalent to each other and are free to move
along the compositional ([001] or [111]) direction.
These atoms will be denoted A$_{1}$ below.
The third A atom, which we denote A$_2$,
lies on the mirror plane between the two B$''$ planes
and is forced by symmetry to remain undisplaced.

We perform local-density approximation
(LDA) calculations on
the above supercells using the Vanderbilt ultrasoft-pseudopotential scheme
\cite{David}. As detailed in Ref.\ \cite{Domenic}, a conjugate-gradient
technique is used to minimize the Kohn-Sham energy functional.
The ultrasoft-pseudopotential approach has two major advantages.
First, it allows for the generation of exceptionally transferable
pseudopotentials because of its use of multiple reference energies during
the construction procedure.
Second, a modest plane-wave cutoff can be used, thus also allowing us to
include the semicore shells in the valence for all the metals considered.
Technically, we chose the plane-wave cutoff to be 25 Ry, which leads to
converged results of physical properties of interest\cite{Domenic}
using roughly 2,700 plane waves per 15-atom supercell.  The Pb
$5d$, Pb $6s$, Pb $6p$, Zr $4s$, Zr $4p$, Zr $4d$, Zr $5s$, Ti
$3s$, Ti $3p$, Ti $3d$, Ti $4s$, Zn $3d$, Zn $4s$, Nb $4s$, Nb
$4p$, Nb $4d$, Nb $5s$, O $2s$ and O $2p$ electrons are treated as
valence electrons in the present study.  Consequently, our
calculations include 122, 134 and 132 electrons per cell in BZN,
PZN and PZT, respectively.  We use the Ceperley-Alder exchange and
correlation \cite{Ceperley} as parameterized by Perdew and Zunger
\cite{Perdew}.  A (6,6,2) Monkhorst-Pack mesh \cite{Monkhorst} was
used for the 1:2$_{[001]}$ structure, corresponding to six k-points
in the irreducible wedge of the Brillouin zone, and providing a
desirable exact mapping onto the (6,6,6) mesh of the primitive
5-atom simple perovskite cell \cite{Domenic}.  A (3,3,3)
Monkhorst-Pack mesh \cite{Monkhorst} was used  for the
1:2$_{[111]}$ structure, corresponding to 10 k-points in the
irreducible zone of the 1:2$_{[111]}$ structure.  We checked that
this mesh leads to converged results by noticing that the total
energy of PZT 1:2$_{[111]}$ structure computed with the (4,4,3) and
the (3,3,3) meshes differed by less than 0.002 eV.

The lattice parameter a$_{0}$, the ``ordered-related''
axial ratio $\xi$, and those atomic displacements consistent with
the symmetries of the ideal structures were fully optimized
by minimizing the total energy and the Hellmann-Feynman forces, the latter
being smaller than 0.04 eV/\AA~ at convergence.

The dynamical effective charge of an atom was determined by
calculating the difference in polarization between the ground state
structure and a structure in which the given atom was displaced slightly
along the compositional direction, the first-order variation of the
polarization with atomic displacement being the dynamical effective charge
Z$^{*}$.
We follow the procedure introduced in
Ref.\ \cite{Domenic2} which consists in directly calculating
the spontaneous polarization as a Berry phase of the Bloch states.
Technically, we use roughly 2,000 Bloch states to assure a good convergence
of the effective charges.

\section{Results:}

\subsection{Energetics}

Figure 2 illustrates the computed energetics for the BZN, PZN, and
PZT cases, showing the energy differences with and without relaxations,
and for 1:2$_{[001]}$ versus 1:2$_{[111]}$ supercells.  The
{\it unrelaxed} structure has an ideal axial ratio $\xi=1$ and
ideal perovskite atomic positions, but a lattice constant
that minimizes the supercell total energy.
On the other hand, the lattice constant, the axial ratio and the
internal coordinates are all optimized in the {\it relaxed} calculations.
One can clearly see that heterovalent and homovalent
alloys differ. In both of the heterovalent compounds BZN and PZN,
the unrelaxed 1:2$_{[111]}$ supercell
is lower in energy by more than 2.5 eV than the unrelaxed
1:2$_{[001]}$ supercell.  In homovalent PZT, on the other hand,
neglecting the relaxations leads to a nearly identical
total energy in the 1:2$_{[001]}$ and 1:2$_{[111]}$ supercells.
This energetic difference between heterovalent and homovalent alloys can
be understood in terms of a model recently developed to explain atomic
ordering in perovskite alloys \cite{LaurentDavid}. In this model, the total
energy of a perovskite with B-site compositional freedom is
approximated by an electrostatic energy given by
\begin{figure}
\epsfxsize=3.0in
\centerline{\epsfbox{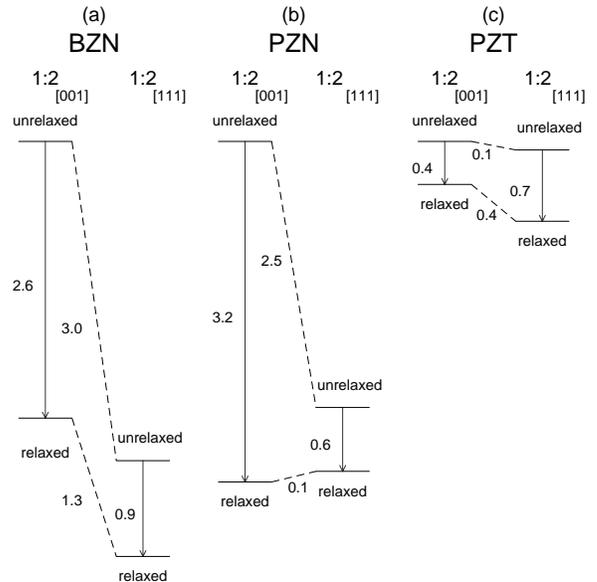}}
\vskip 0.2truein
\caption{Illustration of total-energy differences for (a) BZN, (b) PZN,
and (c) PZT.  The zero of total energy is arbitrarily chosen to be that
of the unrelaxed 1:2$_{[001]}$ structure.}
\end{figure}

\begin{equation}
E = {e^2\over2\epsilon a} \sum_{l\ne l'}
    {\Delta q_l \Delta q_l'\over |\bf l-l'|} \;,
\end{equation}
where $a$ is the cubic lattice constant and $l$ runs over the {\it ideal}
positions of the B atoms (e.g., over Zn and Nb atoms in BZN and PZN, or
Zr and Ti atoms in PZT).
$\epsilon$ is a electronic dielectric constant providing some screening
effects and
$\Delta q_l $ is the relative charge, with respect to the average B charge,
of the B atom in cell $l$. For example, since Zr and Ti belong
to the same column (IV) of the periodic table, $\Delta q_l $ is exactly
zero for any cell in PZT. Applying Eq.\ [1] to PZT then leads
to an energy that is independent of the atomic ordering between Zr and
Ti atoms.  As shown in Fig.\ 2c, this expectation is nearly exactly
demonstrated in our calculations on the
unrelaxed 1:2$_{[001]}$ and 1:2$_{[111]}$ structures.
On the other hand,  in both PZN and BZN,
$\Delta q$ is equal to +1 and $-2$ for Nb and Zn atoms respectively.
Thus, the electrostatic energy $E$ strongly depends on the
atomic ordering of the B atoms,
as demonstrated in Figs. 2a and 2b for the unrelaxed heterovalent
supercells.

One can also evaluate the difference in energy between the {\it unrelaxed}
1:2$_{[001]}$ supercell and the {\it unrelaxed}
1:2$_{[111]}$ supercell for the heterovalent case (BZN or PZN)
using the model of Eq. [1].  The result is
that the unrelaxed 1:2$_{[001]}$ structure is higher in energy
than the unrelaxed 1:2$_{[111]}$ structure by (5.274 a.u.)$/\epsilon a$.
Fitting this formula to the first-principles energetic difference
between the 1:2$_{[111]}$ and the 1:2$_{[001]}$
unrelaxed structures leads to reasonable values of the dielectric constant 
$\epsilon$ (6.3 in BZN and 7.5 in PZN).
Thus, analyzing the first-principles results in the framework of
this electrostatic model suggests that the screening is larger in
Pb-heterovalent than in Ba-heterovalent compounds. This explains why
the energy separation between unrelaxed 1:2$_{[111]}$ and 1:2$_{[001]}$
supercells is larger in BZN than in PZN.

We consider next the relaxation energies (including relaxation
of both atomic and lattice degrees of freedom).
Is is interesting to note that relaxations provide
a comparable decrease in energy (0.6 eV-0.9 eV) in all three
1:2$_{[111]}$ supercells, independently of both the
B-atom valence and the A-atom kind.
On the other hand, the relaxation energy in the 1:2$_{[001]}$
supercell is strongly dependent on chemical effects.
For instance, relaxing atomic and
lattice degrees of freedom in the
Pb-heterovalent alloy PZN leads to a gain
in energy as large as 3.2 eV. This gain is 8 times larger than the
corresponding relaxational decrease in the
Pb-homovalent alloy PZT.
As shown in Fig.\ 2, the different energetic behavior between
Pb-homovalent and Pb-heterovalent
1:2$_{[001]}$ cells has two direct consequences:

(i) The relaxed PZT 1:2$_{[111]}$ is much more
stable (by 0.4 eV) than the relaxed PZT
1:2$_{[001]}$. A similar energetic decrease of
0.3 eV has also been calculated in Ref.\ \cite{Szabo} for
a PZT alloy with 50\% of Zr and 50\% of Ti,  when going
from 1:1$_{[001]}$ to 1:1$_{[111]}$ B-site cation ordering.
(Consistent with our previous notation, the 1:1$_{[001]}$ structure
exhibits a succession of two different B planes along the [001]
direction.)

(ii) The relaxation occurring in PZN leads to a destabilization
of the
1:2$_{[111]}$
supercell with respect to the
1:2$_{[001]}$ one.
The gain in energy by atomic and cell
optimization in the PZN
1:2$_{[001]}$ structure
thus overcomes both the electrostatic
energy difference of 2.5 eV between unrelaxed
1:2$_{[001]}$
and
1:2$_{[111]}$ PZN structures,
and the relaxation energy of 0.6 eV in the PZN
1:2$_{[111]}$ structure \cite{explan}.
This may be consistent with the fact that, unlike many chemically similar
heterovalent alloys, the ground states of
Pb-heterovalent alloys such as PMN and PMT do not
experimentally adopt a 1:2 ordering along a [111] direction.
The destabilization of the
1:2$_{[111]}$
supercell has also been found in PZN
\cite{Wensell}: relaxed calculations performed on
five different order structures predicted that the
1:2$_{[111]}$ supercell was only the
third lowest in energy of those studied.

Relaxation in the 1:2$_{[001]}$ heterovalent BZN supercell also
lowers the energy by a considerable amount (2.6 eV.)
However, in contrast with PZN,  this large decrease is
not able to overcome the very large
electrostatic difference between the unrelaxed
1:2$_{[001]}$ and unrelaxed 1:2$_{[111]}$ BZN
supercells. This could be explained by our above estimates
indicating a smaller dielectric constant in BZN than in PZN.
Our relaxed LDA calculations thus indicate that in BZN the
1:2$_{[111]}$ supercell is significantly lower in energy than the
1:2$_{[001]}$, which is consistent
with the fact that the ground-state structure observed in
BZN (and in many similar Ba-heterovalent alloys)
is the 1:2$_{[111]}$ one \cite{Akbas2,Treiber,Jacobson,Allen,Guo,Kim}.

\subsection{Structural and dielectric effects}

\subsubsection{A-atom effects in heterovalent alloys: BZN {\it vs} PZN}

One may now wonder what are the structural signatures and the dielectric
consequences of the different relaxations that we find in the three
materials under study.  These are summarized in Table I.

First, Table I shows that the heterovalent alloys exhibit similar
trends in lattice parameters. For example, in both BZN and PZN,
the axial ratio is larger for the
1:2$_{[001]}$
structure than for the
1:2$_{[111]}$
one.
In particular, the axial ratio of the
1:2$_{[001]}$ structure
is significantly larger than the ideal value of unity
(by 5\% and 2\% in BZN and PZN, respectively).
This points to the importance
of lattice relaxation in heterovalent alloys.
As usual in LDA calculations, the calculated lattice constants agree
within 1-2\% with the experimental lattice constants of 4.096 \AA~ for
BZN \cite{Didomenico,Nomua} and 4.04 \AA~ for PZN\cite{Bokov}, 
the experimental values having been measured at room temperature
in the cubic paraelectric phase.

Furthermore, we find that all the atomic bonds in the 1:2$_{[111]}$
structure are very similar in the BZN and PZN cases:
(i) the A--O bonds range between 2.80 and 2.86 \AA~ independently
of the A atom (i.e., Pb {\it vs.} Ba);
(ii) the Zn--O bond is lengthened (around 2.05 \AA); and
(iii) the Nb--O bonds are equally divided into a group
with intermediate lengths (1.91 \AA) and a group of longer bonds
with lengths close to that of the long Zn--O bond.
The atomic bonds in the
1:2$_{[001]}$
structure are also qualitatively
similar between PZN and BZN. For example, one may notice four different
features characterizing relaxed heterovalent
1:2$_{[001]}$
structures with respect
to relaxed heterovalent
1:2$_{[111]}$
structures:
(i) four short A$_{1}$--O bonds, $\sim$2.5-2.7 \AA, that are much shorter
than those expected from the sum of ionic radii \cite{Lide} (2.70 and 2.82
\AA~ for Pb--O and Ba--O bonds, respectively);
(ii) four short Zn--O bonds, $\sim$2.0 \AA;
(iii) two very long Zn--O bonds, $\sim$2.2-2.4 \AA, that are
much longer than the 1.95 \AA~expected from ionic radii considerations
\cite{Lide}; and
(iv) one short Nb--O bond of 1.8 \AA.

The differences in atomic bonds between BZN and PZN arise mainly in the
1:2$_{[001]}$ structure. For example, some Zn--O bonds 
are much
longer in BZN 1:2$_{[001]}$ than in PZN 1:2$_{[001]}$; in particular,
the Zn--O bond in BZN 1:2$_{[001]}$ is
$\sim$0.4 \AA~longer than the value of 1.95
\AA~ expected in a purely ionic
perovskite compound \cite{Lide}.  Similarly, some Pb--O bonds in PZN
1:2$_{[001]}$
are shorter by more than 0.1 \AA~ than the shortest Ba--O bonds in BZN
1:2$_{[001]}$.
The formation of very short Pb--O covalent bonds
in PZN, and
the existence of (unstable) anomalous long Zn--O bonds in BZN, are probably
the main quantitative reasons for the larger relaxation energy found
in PZN 1:2$_{[001]}$ compared to BZN
1:2$_{[001]}$.

Table I also shows the Born effective charges $Z^{*}$ for BZN and
PZN supercells, calculated along the compositional direction
(i.e., along either [001] or [111]).
One finds similar trends as in simple perovskite
compounds \cite{Zhong}, namely
values close to +3.0 and +4.0 for Ba and Pb atoms, respectively.
On the other hand, we find that the effective charges of the Nb atoms are
much smaller (by 1-2 units) in the heterovalent alloy than in the
paraelectric states of simple
compounds such as KNbO$_{3}$ and NaNbO$_{3}$ \cite{Zhong}.
Posternak {\it et al.} have shown \cite{Posternak} that
the lengths of Nb--O bonds can have a drastic effect on the
effective charges: they found that the
Z$^{*}$ of Nb decreases by more than 2.5 units when going from the cubic
paraelectric phase to the tetragonal ferroelectric phase of KNbO$_{3}$.
Thus, the fact that Nb atoms have a much smaller effective charge
in BZN and PZN than in simple paraelectric compounds may be attributable
to the Nb--O bond length distribution found in the alloys.
Furthermore, as seen in Table I, the existence of very short Pb--O bonds,
which have been experimentally detected in Pb-heterovalent alloys
\cite{Egami,Chen}, also has a dielectric consequence:
the effective charge of the Pb atom
involved in those Pb--O bonds decreases when going from PZN
1:2$_{[111]}$ to PZN 1:2$_{[001]}$.
A similar decrease of the Pb effective charge
has already been seen when going from a system with ``normal'' Pb--O bonds
to a system with short Pb--O bonds \cite{LaurentJorgeDavid}.
Similarly, heterovalent
1:2$_{[001]}$ structures exhibit both longer Zn--O bonds and
a decrease of the effective charge of the Zn atom.

\subsubsection{Heterovalent effects: PZN {\it vs} PZT}

As shown in Table I and at variance with heterovalent alloys,
the homovalent PZT alloy adopts a near-ideal value of the axial ratio in both
the 1:2$_{[001]}$ and 1:2$_{[111]}$ structures.
The relaxation energy occurring in PZT
1:2$_{[001]}$ is mainly attributable to the existence
of longer Zr--O bonds and shorter
Ti--O bonds, since the Pb--O bonds are very close to their unrelaxed
value of 2.80 \AA.
Unlike the 1:2$_{[001]}$ PZT and the 1:2$_{[111]}$ BZN and PZN
structures, the 1:2$_{[111]}$ PZT structure exhibits short
Pb-O bonds; again, these are presumably responsible for the
stabilization of the relaxed 1:2$_{[111]}$ PZT supercell over
its 1:2$_{[001]}$ counterpart.
From a dielectric point of view, these short Pb--O bonds lead to an
effective charge in PZT 1:2$_{[111]}$ that is smaller than in
PZT 1:2$_{[001]}$.
The main difference in the effective charges of heterovalent
{\it vs.}\ homovalent alloys concerns the
B$'$ atom (i.e., Zn in PZN and BZN {\it vs.} Zr in PZT).
One can see from Table I that Zr has an anomalously large effective
charge $\sim$6.5) with respect to its purely ionic value of +4.
A similar large effective charge of Zr has been found in simple compounds
and results from the weak hybridization of nominally unoccupied
Zr 4$d$ orbitals with
the oxygen 2$p$ orbitals \cite{Zhong}.
On the other hand, since the 3$d$ shell in Zn is {\it below} the
Fermi level, hybridization between the O 2$p$ and Zn 3$d$ orbitals
does not contribute to the effective charge.
(Incidentally, we find that this hybridization is strong,
as reported also in Ref. \cite{Wensell}.)
Hybridization between the O 2$p$ and Zn 4$d$ states could
in principle contribute to $Z^{*}$, but the Zn 4$d$ states are much
higher in energy and therefore unlikely to contribute strongly.
This explains the small values of the Zn effective charges
displayed in Table I.

\section{Conclusions}

We have performed accurate first-principles supercell calculations
to investigate
energetic, structural and dielectric effects in heterovalent and
homovalent perovskite alloys.

Our main findings are that the energetics of the homovalent PZT
alloy is mainly driven by atomic relaxation and
covalent effects. A PZT structure
can thus be stabilized via the formation of short Pb--O bonds. The formation
of these short Pb--O bonds, which have been experimentally seen \cite{Egami},
leads to a decrease of the effective charge of the Pb atoms.

On the other hand, we find that the energetics of the heterovalent
BZN alloys is mainly
driven by electrostatic interactions among the Zn and Nb atoms.
As shown in Ref \cite{LaurentDavid}, these electrostatic
interactions lead to the stabilization of the 1:2 long-range-ordered structure
along the [111] direction. This 1:2 structure does not exhibit any short
A--O bonds, which is consistent with the fact that very short Ba--O bonds
have not been detected (to our knowledge) in heterovalent perovskite alloys.

The situation in PZN is in between those of BZN and PZN, in the sense
that the optimization of the cell parameters and the
formation of short Pb--O bonds in PZN could overcome
the large electrostatic interaction among the Zn and Nb atoms.
The destabilization of the 1:2 structure along the [111] crystallographic
direction can thus happen in PZN. As in PZT, the formation of
these short Pb--O bonds leads to
a decrease of the effective charge of the Pb atoms in PZN.

Finally, one drastic difference in dielectric properties between
heterovalent and homovalent alloys concerns the effective charge of
the larger B-site atom B$'$.  While Zr exhibits anomalous large effective
charge, the Born effective charge
of the Zn atom is very close to its nominal value.

\section{Acknowledgments}

We wish to thank H. Krakauer, M. Wensell, N. Marzari and
A. Garc\'{\i}a for useful discussions.
This work is supported by the ONR grant N00014-97-1-0048.
Cray C90 computer time was provided at the NAVO MSRC DoD HPC center.

\begin{table}[h]
\widetext
\caption{Lattice parameters $a_{0}$ and $\xi$ (see text), atomic distances
$d$, and effective charges $Z^{*}$ of
Ba(Zn$_{1/3}$Nb$_{2/3}$)O$_{3}$ (BZN),
Pb(Zn$_{1/3}$Nb$_{2/3}$)O$_{3}$ (PZN), and
Pb(Zr$_{1/3}$Ti$_{2/3}$)O$_{3}$ (PZT) supercells.
A$_{1}$ is an A atom that relaxes along the compositional direction,
while A$_{2}$ is prevented from relaxing from its ideal position by
symmetry.
B$'$ is the larger B atom (Zr in PZT, Zn in BZN and PZN);
B$''$ is the smaller one (Ti in PZT, Nb in BZN and PZN.)
For each kind of bond, the number of them occuring per
supercell is indicated in brackets.
Our calculations agree with the results of Ref.\ \protect\cite{Wensell}
within 0.04 \AA~ for all the bond lengths of the PZN
1:2$_{[111]}$ structure.}
\medskip
\label{Table I}
\begin{tabular}{ldddddd}
& \multicolumn{2}{c}{BZN} & \multicolumn{2}{c}{PZN} & \multicolumn{2}{c}{PZT} \\
 &~~1:2$_{[001]}$ &~~1:2$_{[111]}$ &~~1:2$_{[001]}$ &~~1:2$_{[111]}$
 &~~1:2$_{[001]}$ &~~1:2$_{[111]}$
 \\
\tableline
 $a_{0}$ (\AA) & 4.02 & 4.02  &3.97 &4.02 &3.97 &3.97 \\
 $\xi$  & 1.05 & 1.00  &1.02 &0.98 &1.00 &1.00 \\
 d(A$_{1}$--O) (\AA)
  & 2.66 [4] & 2.85 [3] & 2.54 [4] & 2.80 [3] & 2.74 [4] & 2.60 [3] \\
  & 2.91 [4] & 2.85 [3] & 2.87 [4] & 2.82 [3] & 2.81 [4] & 2.81 [6] \\
  & 3.28 [4] & 2.86 [6] & 3.28 [4] & 2.85 [6] & 2.91 [4] & 3.07 [3] \\
 d(A$_{2}$--O) (\AA)
  & 2.81 [8] & 2.83 [6] & 2.74 [8] & 2.80 [6] & 2.76 [8] & 2.76 [6] \\
  & 2.84 [4] & 2.84 [6] & 2.81 [4] & 2.85 [6] & 2.81 [4] & 2.81 [6] \\
 d(B$'$--O) (\AA)
  & 2.01 [4] & 2.07 [6] & 1.98 [4] & 2.04 [6] & 1.98 [4] & 2.06 [6] \\
  & 2.38 [2] &          & 2.19 [2] &          & 2.11 [2] &          \\
 d(B$''$--O) (\AA)
  & 1.82 [1] & 1.91 [3] & 1.84 [1] & 1.91 [3] & 1.89 [1] & 1.98 [3] \\
  & 2.01 [4] & 2.06 [3] & 1.99 [4] & 2.05 [3] & 1.93 [1] & 2.00 [3] \\
  & 2.10 [1] &          & 2.06 [1] &          & 1.98 [4] &          \\
 Z$^{*}$(A$_{1}$)  & 3.10 & 3.08  &3.48 &4.44 &4.04 &3.78 \\
 Z$^{*}$(A$_{2}$)  & 2.49 & 2.20  &3.43 &3.34 &3.53 &3.41 \\
 Z$^{*}$(B$'$)  & 2.24 & 3.16  &2.59 &2.95 &6.69 &7.76 \\
 Z$^{*}$(B$''$)  & 6.88 & 6.54  &7.82 &6.45 &6.65 &5.81 \\
\end{tabular}
\narrowtext
\end{table}

\end{document}